\documentclass[letterpaper]{article}
\usepackage{isea}
\usepackage[pdftex]{graphicx}
\usepackage{times}
\usepackage{amsmath}
\usepackage{helvet}
\usepackage{courier}
\usepackage[numbers]{natbib}
\title{Can AI Recognize the Style of Art?\\ 
Analyzing Aesthetics through the Lens of Style Transfer}
\author{Yunha Yeo\\
Graduate School of Culture Technology, KAIST\\
Daejeon, South Korea\\
yunhayeo@kaist.ac.kr\\
\newline
\newline
\And
Daeho Um\thanks{Corresponding author}
\\
Samsung Advanced Institute of Technology (SAIT)\\
Suwon, South Korea\\
daeho.um@samsung.com\\
}

\begin{document} 

\maketitle

\begin{abstract}

This study investigates how artificial intelligence (AI) recognizes style through style transfer—an AI technique that generates a new image by applying the style of one image to another. Despite the considerable interest that style transfer has garnered among researchers, most efforts have focused on enhancing the quality of output images through advanced AI algorithms. In this paper, we approach style transfer from an aesthetic perspective, thereby bridging AI techniques and aesthetics. We analyze two style transfer algorithms: one based on convolutional neural networks (CNNs) and the other utilizing recent Transformer models. By comparing the images produced by each, we explore the elements that constitute the style of artworks through an aesthetic analysis of the style transfer results. We then elucidate the limitations of current style transfer techniques. Based on these limitations, we propose potential directions for future research on style transfer techniques.


\end{abstract}


style transfer, AI-generated art, AI aesthetics, computer vision

\section{Introduction}


Artificial intelligence (AI), which aims to imitate human intelligence, has achieved significant advancements across various fields, such as medicine, agriculture, and education.~\cite{hamet2017artificial, holmes2004artificial,jha2019comprehensive,hwang2020vision, chen2020artificial} Surprisingly, AI has even extended its applicability to the field of art by creating artworks, once considered a unique human domain. For example, AI demonstrated its artistic ability by winning an award in the Lumen Prize with an AI-generated artwork titled \textit{Portrait of Edmond de Belamy} in 2018. Recently, various AI models have been closely related to art, and style transfer is one of them. Style transfer is a computer vision technique that interprets artworks through ``style", a concept traditionally discussed by humans. Specifically, style transfer methods apply the style of one image (\textit{i.e.}, the style image) to the content of another image (\textit{i.e.}, the content image), resulting in a new synthetic image (\textit{i.e.}, the output image).~\cite{jing2019neural}


Style transfer technologies have been developed by leveraging cutting-edge computer vision models, all of which share a fundamental mechanism. Initial style transfer models utilize convolutional neural networks.~\cite{li2016combining, wang2017multimodal, risser2017stable} Due to their ability to replicate the style of paintings, style transfer methods have garnered significant attention, leading to the release of numerous publicly available codes and widespread use. On the other hand, ChatGPT, which employs a high-performance Transformer model, has recently demonstrated its effectiveness in generating high-quality artistic outputs and has gained popularity.~\cite{vaswani2017attention}

\begin{figure}[t]
\includegraphics[width=\linewidth]{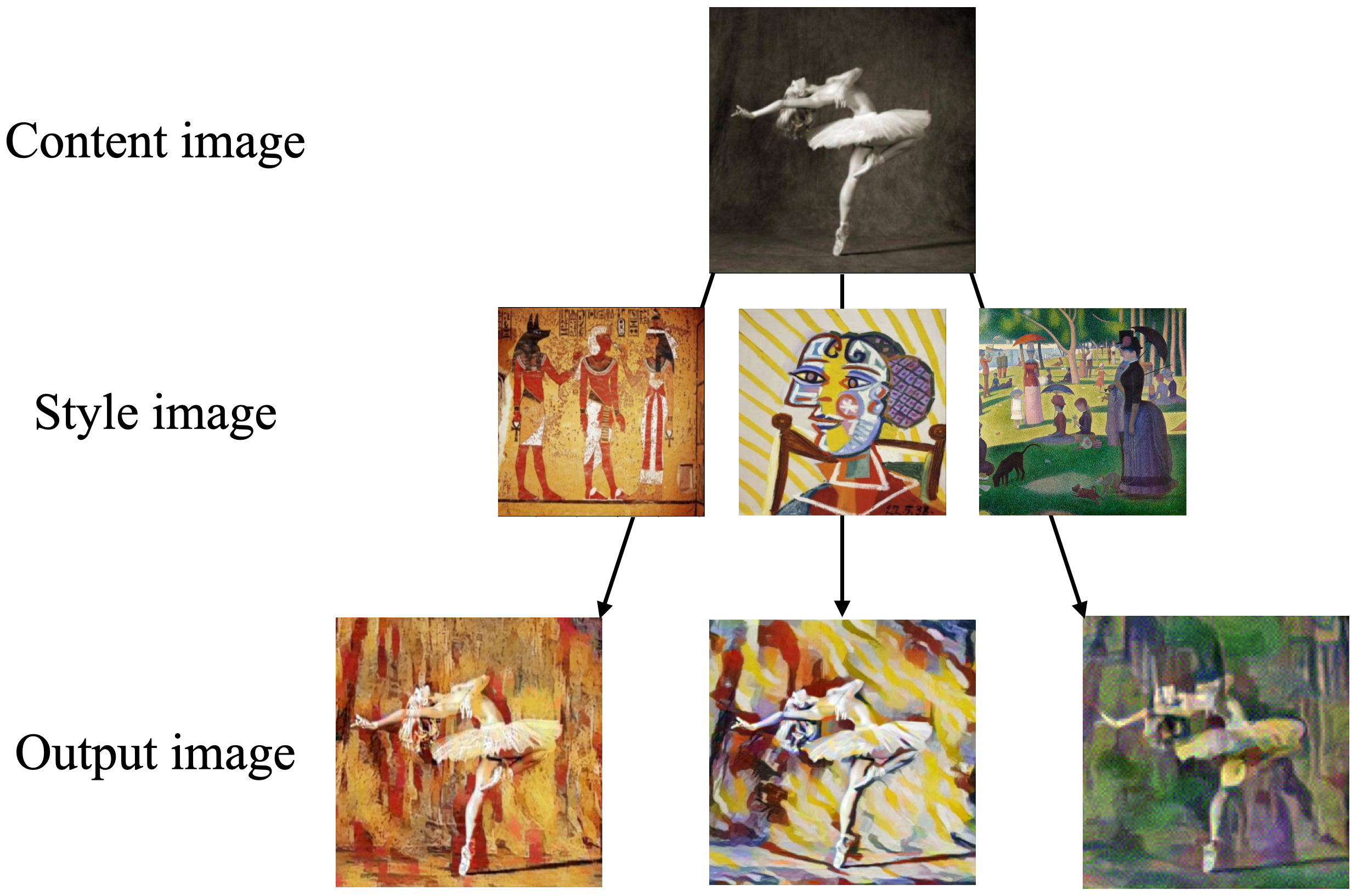}
\caption{Results of CNN-based style transfer applied to a ballerina image as the content image, with various famous paintings used as style images.}
\vspace{-3mm}
\label{fig:teaser}
\end{figure}


As illustrated in Figure~\ref{fig:teaser}, existing style transfer models demonstrate their ability to stylize images. Despite their potential, neither CNN-based nor Transformer-based style transfer models are fully capable of capturing the complete style of artworks. Both models primarily focus on simple visual elements of the style images, such as color and texture. However, style is a highly intricate and multifaceted concept that encompasses not only these simple elements but also the artist's intention and the deeper meanings embedded within the artwork. Nevertheless, discussions around style transfer have largely centered on its technological aspects, and the performance of style transfer models has predominantly been assessed qualitatively, with no clear or consistent evaluation criteria established. Furthermore, given that style transfer is deeply intertwined with visual art, its evaluation should also consider complex dimensions such as artistic integrity.

In this paper, we present a pioneering analysis of how AI models perceive and interpret artistic styles. Specifically, we identify the stylistic elements recognized by CNN-based and Transformer-based style transfer models and compare their respective capabilities. Furthermore, we reassess whether current AI models can genuinely capture and transfer style in a manner that aligns with human perception, thereby revealing critical limitations and suggesting potential directions for future advancements in style transfer models.



The main contributions of this paper are threefold. Firstly, to the best of our knowledge, this work is the first attempt to analyze style in terms of image elements from the perspective of AI. Secondly, this paper reevaluates whether current AI models genuinely capture and transfer style in a manner consistent with human understanding, highlighting key limitations of existing style transfer models. Lastly, we offer guidance for future research in the field of style transfer, encouraging its expansion into the realm of human perception.

\section{Related work}

\vspace{-1mm}

\subsection{Style}

The word `style' originates from a Roman writing implement, \textit{stilus}.~\cite{andrews1896latin}
Various methodologies (\textit{e.g.}, formalism, iconography, psychoanalytic) and schools of thought (\textit{e.g.}, structuralism, feminism, Marxism) exist in art history research, which aims to synthesize the style and content of artworks with the artistic systems and societal aspects of the times in which the artworks were created. For example, Heinrich Wölfflin established a style theory related to modern trends by comparing and analyzing artworks from various periods.~\cite{wolfflin1950principles} Wölfflin identified common elements across individual styles, period styles, and national styles, leading to the classification of styles according to historical changes.

On the other hand, Erwin Panofsky defined style as the expression of the philosophical, cultural, and social values of a specific time period. Panofsky argued that style encompasses the spiritual and ideological essence of an era, which is visually manifested in the artwork. Thus, Panofsky, when analyzing style, considered not only aesthetic elements but also the historical and social contexts in which the artwork was created.~\cite{panofsky1967studies} Another art historian, Ernst Gombrich asserted that style is not merely the representation of the outer world, but also a mental image, reflecting the invisible inner world. In other words, Gombrich claimed that since creators express what exists in their mind through their works, it is impossible to classify works into fixed categories using an objective reference point.~\cite{gombrich2023art} Additionally, Gombrich suggested that images should be analyzed from a psychological perspective, focusing on the factors that influence the creative process.~\cite{gombrich2023art} While there have been diverse perspectives and interpretations of styles, no fixed standard exists. In this work, we attempt to analyze style by utilizing AI's perspective, \textit{i.e.}, how style is interpreted by AI. 

\vspace{-1mm}

\subsection{Style Transfer}

Style transfer is a popular topic in computer vision, and many researchers have focused on enhancing style transfer methods from a technical standpoint. With the rapid development of neural networks, which have ushered in the AI era, a CNN-based approach pioneered style transfer using neural networks.~\cite{gatys2016image} Building on this work, various CNN-based style transfer methods have been proposed, leading to significant improvements in the quality of output images.~\cite{li2016combining, wang2017multimodal, risser2017stable} Recently, as Transformer models have proven to be more effective in a variety of computer vision tasks, Transformer-based models have also been developed for style transfer.~\cite{vaswani2017attention,deng2022stytr2} These Transformer-based techniques have achieved notable technical advancements, and some of these techniques have been adopted in systems like ChatGPT. While many related studies have contributed to the technical progress of style transfer, to the best of our knowledge, no attempt has been made to integrate AI-driven style transfer with visual art theory, which focuses on analyzing the aesthetics of the generated images.

\section{Mechanism of Style Transfer}
\label{sec:mechanism}


The process of style transfer, which requires two types of input, can be divided into two phases: (1) the conversion from content image to output image, and (2) the conversion from style image to output image. In this section, we compare and contrast the mechanisms of CNN-based style transfer and Transformer-based style transfer. Understanding the distinct characteristics of these two mechanisms is important as it highlights the core differences in how each AI model perceives style.

\subsection{CNN-Based Style Transfer}

\begin{figure*}[h]
\includegraphics[width=\linewidth]{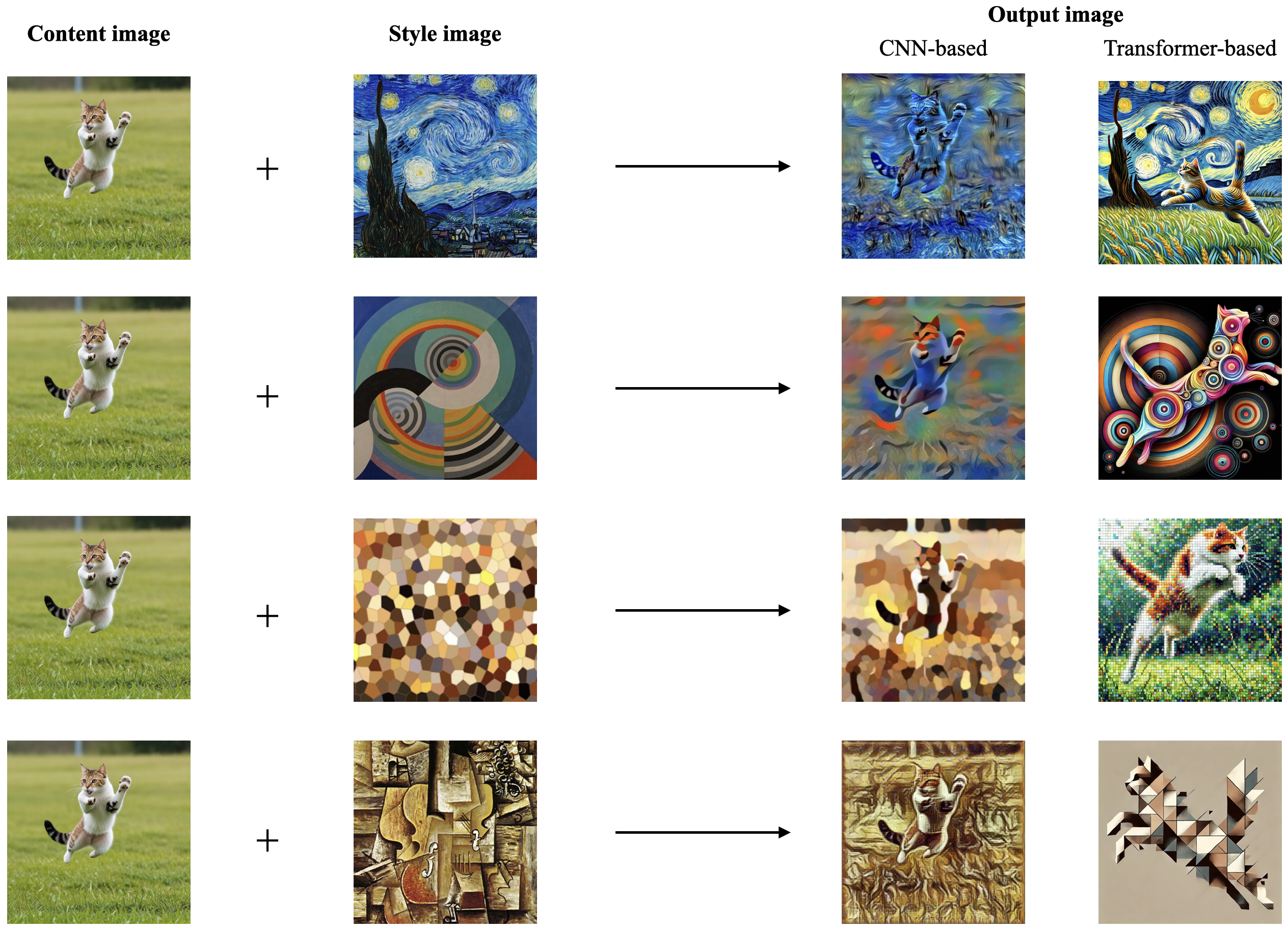}
\caption{Comparison of CNN-based and Transformer-based style transfer applied to a cat content image with various artworks as the style images.}
\label{fig:catcompare}
\end{figure*}

First, we explain the mechanism of CNN-based style transfer. CNN consists of deep layers, where each layer generates a feature map as an intermediate output before producing the final output.~\cite{li2021survey} Training CNN for style transfer has two objectives: to minimize the difference between the final image and the feature map of the input content image and to ensure that the texture of the final image resembles that of the input style image.~\cite{jing2019neural} To achieve these objectives, a loss function used for the CNN training is the sum of a content loss function and a style loss function, each designed for the respective objectives. The training is done by iteratively minimizing the total loss function using training images.

First, the content loss function is calculated by measuring the error between the feature maps of the content image and those of the output image. This error is computed on a per-location basis, preserving local information. In other words, minimizing the content loss function ensures that pixels in corresponding locations of the content and output images become similar. Second, the style loss is computed by minimizing the error between the Gram matrix of the style image and that of the output image. Specifically, the Gram matrix computed using multiple feature maps represents the relationships between the feature maps. This Gram matrix is known to reflect the style of the produced image. Therefore, minimizing the style loss function causes the output image to adopt a style similar to that of the style image.

Figure~\ref{fig:catcompare} illustrates examples of style transfer, where the cat in the content image is stylized with the style of the style image. Thus, the loss function for style transfer, which is the sum of the content loss function and the style loss function, helps to find the optimal weights that satisfy both objectives mentioned above.

\begin{figure*}[t]
\includegraphics[width=\linewidth]{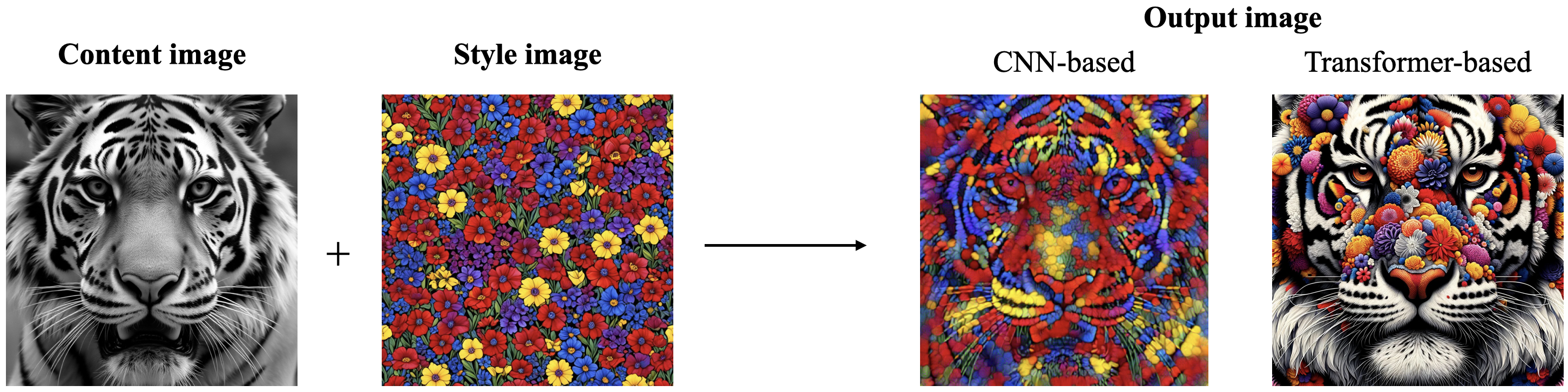}
\caption{Comparison of CNN-based and Transformer-based style transfer when using a tiger image as the content image.
}
\label{fig:tigercompare}
\end{figure*}


\subsection{Transformer-Based Style Transfer}

While Transformer-based style transfer and CNN-based style transfer share the core mechanism that minimizes the total loss, the sum of the content loss and the style loss, a key difference lies in the self-attention mechanism used by Transformer.~\cite{deng2022stytr2} Unlike CNNs, which rely on convolutional layers to focus on local pixel groups, Transformers leverage the self-attention to process global dependencies across the entire image.~\cite{vaswani2017attention} This means that every pixel in the content image can interact with all other pixels, allowing the model to capture both local and global contexts in a single pass.

The self-attention mechanism works by computing attention scores between all pairs of image patches, assigning greater importance to relevant regions and stylistic features from both the content and style images. Each attention layer transforms the content image by blending information from other areas of the image, which allows the model to apply stylistic changes more flexibly and globally. As a result, the content image's semantic structure is preserved while style elements—such as color, texture, and patterns—are globally redistributed across the entire output image.

Additionally, Transformer models operate by splitting an image into patches that are processed in parallel. Each patch is encoded into a feature representation that is influenced by other patches via the self-attention mechanism. This process ensures that both the local details and the global structure of the image are captured and transformed according to the style image. The ability of Transformers to process the entire image at once, rather than incrementally like CNNs, enables them to better capture complex and global stylistic patterns, making Transformers highly effective for more abstract or large-scale style transfer tasks.

\begin{figure*}[t]
\includegraphics[width=\linewidth]{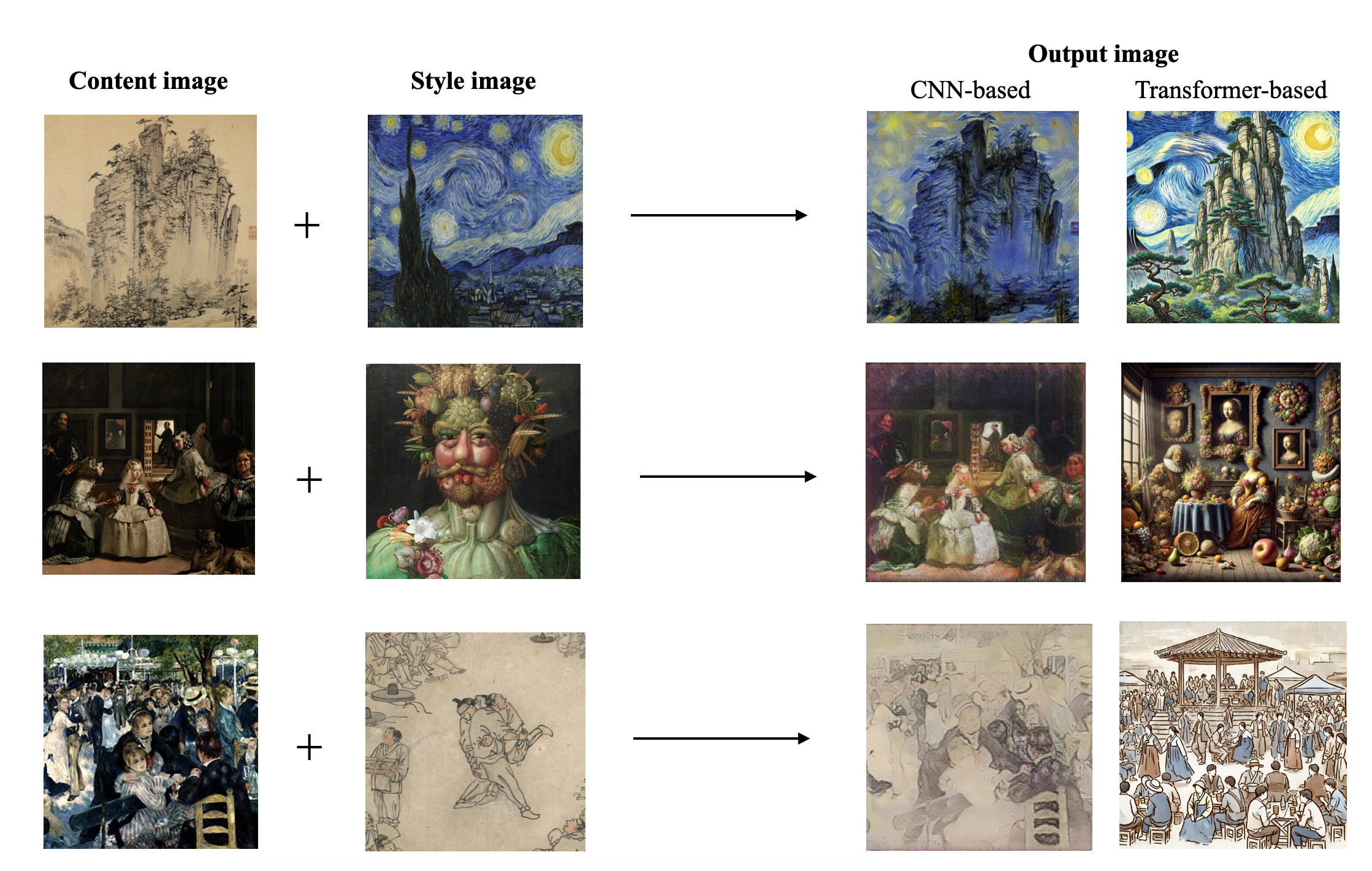}
\caption{Comparison of CNN-based and Transformer-based style transfer.}
\label{fig:example}
\end{figure*}

\section{Comparison of CNN-Based Style Transfer and Transformer-based Style Transfer} 

In this section, we analyze the similarities and differences between CNN-based style transfer and Transformer-based style transfer. We focus on how the two style transfers preserve the subject from the content image and the style from the style image. Style transfer consists of two stages of processing: the content-to-output conversion process and the style-to-output conversion. In the content-to-output conversion process, the subject in the content image is preserved. These overall processes raise questions about how the style of an artwork is interpreted by AI and how viewers perceive that the style is transferred.


\subsection{CNN-Based Style Transfer}

The content loss function is computed as the total difference between the feature map values at corresponding locations in the content image and the output image. Thus, CNNs focus on the local context of the content image and aims to preserve its semantic elements, preserving the boundaries and spatial arrangement of the content image.

The boundaries of the content are preserved with the help of the content loss function. Furthermore, the smaller the content loss function, the more similar the feature map values are between the content image and the output image at corresponding locations. For example, if the content image represents a tiger, as shown in Figure~\ref{fig:tigercompare}, the boundaries corresponding to the tiger's eyes, nose, and facial contours are preserved in the output image.

The conversion from the content image to the output image not only preserves the boundaries but also the spatial arrangement. For instance, in Figure ~\ref{fig:tigercompare}, if the lower center area of the content image shows the lion's mouth, then the lower center area of the output image will also display the lion's mouth. This is because convolution, the central operation of CNN, is inherently designed to preserve spatial relationships. As a result of this characteristic of convolution, the spatial arrangement of the content image is retained in the feature map. Therefore, convolution ensures the feature maps of both images have similar values at corresponding locations implying that the pixel values in the same regions of the two images will also be similar. The preservation of locality is also applied to the style loss function. 

\subsection{Transformer-Based Style Transfer}

As seen in Figure~\ref{fig:tigercompare}, Transformer-based style transfer also tends to preserve the boundaries and spatial arrangement of the content image. However, unlike CNNs, Transformer models do not focus solely on the local context and neighboring pixels of the image; instead, they incorporate information from the entire image. In other words, Transformer-based style transfer is capable of handling global context. Due to this characteristic, Transformer-based style transfer can interpret the underlying intention embedded within certain styles. In addition, while CNNs focus heavily on local features to make stylistic adjustments, Transformers can process the global context of an image due to their attention mechanism. This enables Transformer-based models to better understand and manipulate long-range dependencies between different parts of an image, which can lead to generating stylistically cohesive images, especially when dealing with complex textures or global stylistic patterns.

Transformers are designed to process images as a series of patches, each containing both content and stylistic information. These patches are processed simultaneously, allowing the Transformer-based model to apply stylistic changes that impact the image on a global scale, rather than focusing solely on small regions. As a result, the entire spatial arrangement of the content image is preserved, but the colors, textures, and patterns from the style image are applied more flexibly compared to CNNs. This method enables the Transformer-based model to manipulate the overall style of the image, including complex stylistic features that span across the entire image. 

Furthermore, when Transformer-based style transfer aims to capture the intention embedded in the style, it may alter the boundaries, spatial arrangement, and even the shape of the content image, creating a transformed image that retains both content and stylistic intent. Figure~\ref{fig:catcompare} illustrates this unique characteristic of Transformer-based style transfer, highlighting the significant differences between the outputs of Transformer-based and CNN-based style transfer.

\section{Analyzing Style Elements through Style Transfer Models}


In this section, through extensive experiments using CNN-based and Transformer-based models, we investigate which visual elements each style transfer model recognizes as style. Additionally, we examine the extent to which each model understands the style of artworks, moving beyond discussing simple visual elements.




\begin{figure*}[t]
\includegraphics[width=\linewidth]{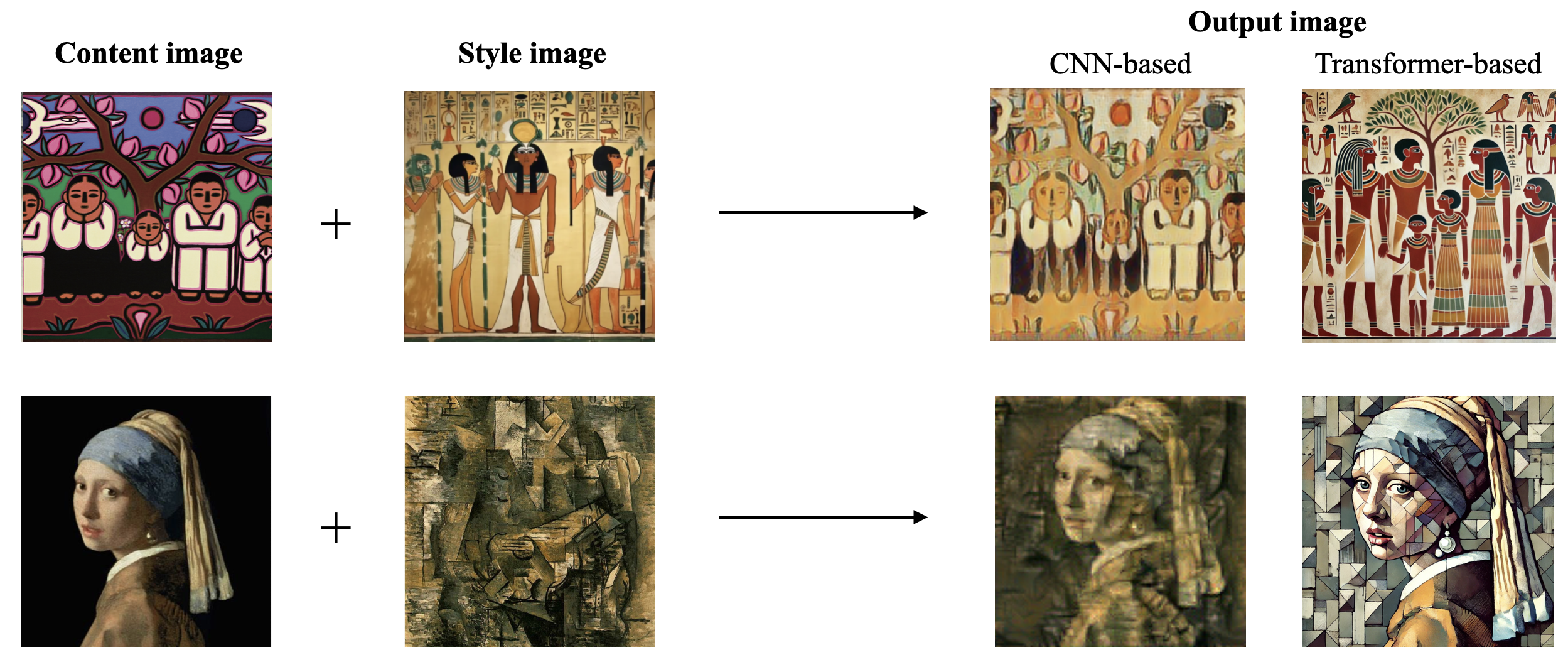}
\caption{Examples that show the limitations of existing style transfer models.}
\label{fig:limitcompare}
\end{figure*}

\subsection{Emotional Tone}

Clive Bell, an English art critic, introduced the concept of ``significant form" as the expression of a unique emotion felt for reality.~\cite{bell2018aesthetic} Bell argued that formal structural elements can alone have the power to evoke emotional responses from viewers and that once this occurs, the object can truly become an artwork. This suggests that transferring structural elements of the style image, such as lines and texture, can convey the emotional tone associated with the style image. The top row of Figure~\ref{fig:example} demonstrates the results of style transfer using \textit{The Starry Night} painted by Vincent van Gogh as the style image. Van Gogh’s use of impasto in \textit{The Starry Night}, where thick layers of paint create a three-dimensional texture, elicits a vibrant and intense emotional tone. The unique texture and parallel lines of \textit{The Starry Night} evoke a dynamic and dreamlike emotional tone.

In the case of the CNN-based model, as shown in the figure, colors are transferred faithfully, preserving the style image's color combinations. In addition, the use of parallel lines and the unique texture is also successfully maintained. 
Since visual factors significantly influence emotional tone, the faithful transfer of color and structural elements allows the output images to reproduce the emotional impact of the style image.

On the other hand, we can observe that the Transformer-based model captures visual elements more holistically and globally compared to the CNN-based model. As shown in Figure~\ref{fig:example}, the Transformer-based model does not only take color and texture into account but also the overall composition and visual patterns. Considering diverse visual elements, the Transformer-based model has a deeper understanding of both the context and style images. Thus, the Transformer-based model can facilitate more creative and deeper transformations of images, delivering a more comprehensive emotional tone. In a nutshell, while the CNN-based and Transformer-based models commonly recognize fundamental structural elements, such as color-related features and texture, the Transformer-based model further captures global visual elements.

\subsection{Signs} A sign is the fundamental unit in semiotics, representing an entity beyond itself. Connotation and culture are two key concepts related to signs. While connotation refers to the process by which artists embed intrinsic meanings within their artworks, the culture of the scene depicted in an artwork is often conveyed through elements that reflect local or historical characteristics. Connotation and culture are expressed through signs, enabling viewers to interpret inherent meanings. Thus, when common signs are present in two different artworks, the style of those artworks is often perceived as similar, implying that signs are an important style element.






\subsubsection{Connotation} 

Signs often do not directly reveal meanings; instead, they express themselves in suggestive and symbolic ways. Roland Barthes referred to this characteristic of signs as connotation.~\cite{barthes1999rhetoric} Connotations in images often involve multiple layers of meaning, including subjective and implicit ideas. In traditional paintings, specific colors or objects imply cultural or historical meanings that transcend superficial elements, offering profound interpretations. The middle row of Figure~\ref{fig:example} shows the results when Diego Velázquez's \textit{Las Meninas} is used as the content image and Giuseppe Arcimboldo’s \textit{Vertumnus} as the style image. In \textit{Vertumnus}, Arcimboldo used fruits, vegetables, and other natural elements to compose a face. Each part of the face consists of seasonal produce, symbolizing the god Vertumnus’s connection to the cycles of nature. This painting is also a portrait of Emperor Rudolf II of the Holy Roman Empire as Vertumnus, and the natural elements serve as an imperial allegory that symbolically links Emperor Rudolf II to the cycles and abundance of nature. As shown in the figure, the CNN-based model fails to convey these signs with connotative meanings from the style image to the output image.  In contrast, the Transformer-based style transfer model reallocates the forms of vegetables and fruits in the new content image, demonstrating its ability to convey the signs more insightfully.



\subsubsection{Culture}

Roland Barthes further argued that an image is not a simple representation but a complex layering of meanings, including cultural and social implications.~\cite{barthes1999rhetoric} This concept of the image, as described by Barthes, can be applied to the analysis of image styles, where style is understood as a form of cultural semiotics. \textit{Ssireum} drawn by Kim Hong-do is a painting illustrating people gathering around to watch a Korean wrestling match, which reflects the lifestyle and leisure activity of Korea in the late 18th century. The bottom row of Figure~\ref{fig:example} presents the results when Pierre-Auguste Renoir's \textit{Bal du moulin de la Galette} is used as the content image and \textit{Ssireum} as the style image. In \textit{Ssireum}, we can observe traditional Korean customs, which serve as signs representing the culture. However, as shown in Figure~\ref{fig:example}, the CNN-based model falls short in conveying these cultural elements. In contrast, the Transformer-based model successfully transfers them, making the output image as if it were a snapshot of Korean life in the late 18th century.

 \section{Limitations of Existing Style Transfer Techniques}


In this section, we examine the limitations of current style transfer models from an aesthetic perspective. Figure~\ref{fig:limitcompare} exhibits the examples that reveal the two types of style transfer models' clear limitations. 

The CNN-based style transfer model effectively captures surface-level attributes such as color and texture but face challenges comprehending deeper artistic concepts, particularly in complex styles like ancient Egyptian art and Cubism. For example, the result of the output image in the top row of Figure~\ref{fig:limitcompare} is unsatisfactory when an ancient Egyptian style is applied to a painting in the Joseon Dynasty, a period in Korean history. The output image fails to replicate the unique features of Egyptian art, characterized by a combination of side and front views. Although the colors and textures were transferred, the essential elements of Egyptian style—such as its symbolic portrayal of the human figure and the religious and spiritual meanings related to the afterlife—are not conveyed in the context of the Joseon painting.~\cite{hartwig2014style} Similarly, in the bottom row of Figure~\ref{fig:limitcompare}, where Georges Braque’s \textit{The Portuguese} style is being transferred, the CNN-based model fails to capture Cubism's core idea of multi-perspective composition and spatial deconstruction. While the output image preserves color and texture from the style image to some extent, it loses the essence of Cubism, which focuses on depicting objects from various angles. In summary, the intention of the artist and the process of making the style are not comprehended by these AI models.

On the other hand, the Transformer-based style transfer model demonstrates better performance compared to the CNN-based style transfer model in interpreting the underlying intentions of certain styles, particularly concerning the global structure and composition of the image. For example, in Figure~\ref{fig:limitcompare}, the Transformer-based model shows a deeper understanding of ancient Egyptian art by capturing its compositional characteristics. However, even the Transformer-based model struggles with Cubism, since it fails to reproduce the multi-dimensional perspectives and abstract forms central to the movement. While the Transformer-based model processes style with a more holistic approach, it still lacks the intuitive understanding of complex styles. Although style transfer models have developed the ability to interpret style at academic and technical levels, they still face challenges in conveying abstract concepts and artistic intentions, which are integral to styles like Cubism. For instance, in Cubism, Georges Braque and Pablo Picasso depicted objects from multiple perspectives, integrating fragmented perceptions into their work. The flattening of depth in Cubist art was not intended to create a flat image but was instead the result of exploring various angles and perspectives. Current AI models, however, are unable to capture the intention behind this process-driven approach, reducing the outcome to a mere visual byproduct rather than a reflection of the artist’s method.


Unlike humans, who can intuitively and emotionally engage with artworks, existing style transfer models are still incapable of grasping the deeper essence of art and the underlying concepts that define its creation. Further supporting this limitation is Ernst Gombrich’s theory, which posits that style is not merely a representation of the external world but also a reflection of a mental image. While the CNN and Transformer-based models can replicate outer appearances, they are unable to transfer the internal vision or concept that the artist seeks to express. The mental image, which encompasses the deeper intent and subjective interpretation behind the artworks, remains beyond the reach of current AI models. Ultimately, AI models, rooted in pattern recognition and replication, struggle to understand and generate multi-dimensional perspectives involving complex artistic concepts. These limitations highlight the gap between AI’s ability to mimic surface-level aesthetics and the profound, often philosophical understanding of art that involves engaging with both form and content as well as the deeper artistic intentions behind them.


\section{Conclusion}

With the question ``Can AI recognize style?" in mind, we compared and contrasted two style transfer models that operate through different mechanisms. The early style transfer model using CNN captured the superficial visual elements, such as color and texture in images. On the other hand, the recent Transformer-based model demonstrated the ability to grasp global context and explore deeper aspects of style. While the existing style transfer models showed their potential for understanding and analyzing style, they still fell short of capturing the essence of art that lies beyond superficial traits from an aesthetic perspective. We also conducted an in-depth analysis of visual components constituting an image's style through the lens of AI. This paper shows that aesthetics analysis can provide valuable insights for AI models by revealing their limitations. We suggest that future AI style transfer models should advance toward seeking aesthetic values rather than merely mimicking surface-level traits.
\bibliographystyle{isea}
\bibliography{iseabib}

\section{Author Biographies}

Yunha Yeo received the B.F.A. degree from the School of the Art Institute of Chicago, Chicago, Illinois, in 2018, and the M.A. degree in Visual Culture Studies from Korea University, Seoul, South Korea, in 2024. She is currently pursuing the M.S. degree at the Graduate School of Cultural Technology, Korea Advanced Institute of Science and Technology, Daejeon, South Korea. Her current research interests include AI-generated art, digital humanities, and media studies.

Daeho Um received the B.S. degree in
electrical engineering from Korea University, Seoul, 
South Korea, in 2019, and the Ph.D. degree in electrical and computer engineering with Seoul National University,
Seoul, South Korea, in 2024. He is currently a Staff
Engineer with Samsung Electronics, Kyeong-Gi, South Korea. His current research interests
include graph-based machine learning, data imputation, and computer vision.


\end{document}